\documentclass[12pt]{article}       
                         
\title{Cosmic Topology : Twenty Years After}
 \author{Jean-Pierre Luminet, Laboratoire Univers et Th\'eories\\ Observatoire de Paris-CNRS-Universit\'e Paris Diderot (France)\\email : jean-pierre.luminet@obspm.fr}

\begin{document}

\maketitle

\begin{abstract}What is the shape of the Universe? Is it finite or infinite ? Is space multi-connected to create ghost images of faraway cosmic sources? After a ``dark age'' period, the field of cosmic topology has now become one of the major concerns in astronomy and cosmology, not only from theorists but also from observational astronomers. Here I give a personal account of the spectacular progress in the field since the beginning of the 1990's, when I started to work in it. 

\end{abstract}

\bigskip

\section{Back to the future}
\label{sec:1}

It is a real pleasure for me to contribute to the celebration of the $75^{th}$ birthday of my friend Andrey Grib. 
I met him for the first time exactly twenty years ago, when he invited me as a lecturer at the Second Alexander 
Friedmann International Seminar on Gravitation and Cosmology that took place in September 12--19, 1993 in 
St. Petersburg, a conference of which Andrey was a chairman and co--organizer.

For me it was the opportunity to give my first seminar about a subject that I was just beginning to investigate, 
namely multi--connected cosmologies~\cite{lum94}. The Friedmann Seminar was a perfect place for this, 
since the Russian pioneer of modern relativistic cosmology was also a pioneer in ``cosmic topology'' --  
a term later coined in 1995 by myself in a large review paper carrying this title~\cite{lalu}.  

As it is well-known, the discoveries of non--static universe models by
Friedmann in 1922--1924, then by Lema\^{\i}tre in 1927--1931, have definitively changed the
face of cosmology. What is much less known is that their seminal articles also contained
deep discussions of the global structure of the Universe. Friedmann quickly took notice
of the incompleteness of the  general relativity theory concerning the topological
question. The discussion that he gave can be found in his article from 1924
where he discovered eternally expanding hyperbolic models for the universe~\cite{Fri24}. In
the final paragraph, the Russian scientist clearly defined the fundamental limitations of the cosmological
theory founded on general relativity: ``Einstein's world equations, without
additional assumptions, do not allow us to decide the question of the finiteness
of our world.'' He undertook to define how space could become finite if one
identified different points (which, in topological language, renders the space multi--connected). 
He also foresaw how this possibility allowed for the existence of
phantom sources, in the sense that at a single point an object coexists with its multiple images.
``A space of positive curvature is always finite,'' he added, but mathematical
knowledge ``does not allow us to settle the question of the finiteness of a space of
negative curvature.''

In contrast to Einstein, Friedmann had no prejudice in favor of a simply-connected
topology. Nevertheless, he believed, following the example of most physicists of the time, 
that only spaces of finite volume were physically admissible for
describing real space. The universe models proposed by Einstein and de Sitter in
1917, next by Friedmann in 1922, all had a positive spatial curvature and therefore
satisfied the criterion of finiteness. With models of negative curvature, the problem
was more arduous: hyperbolic space, in its simplest version (with a simply-connected topology), 
extends to infinity. Therefore, even at the moment of ``creation''
of the Universe, space was itself infinite; in other words, the Universe was
not born as a point, rather the expansion started at every point of a preexisting
infinite space. Aware of this difficulty, Friedmann saw a loophole in the fact that
Einstein's equations were not sufficient for deciding if space was finite or infinite,
even if the curvature is negative or zero. One must make additional hypotheses to
specify the conditions at the limits, in particular, whether certain points in space
are identified with others or not. The whole problem of cosmic topology was thus
posed, but Friedmann did not have sufficient mathematical tools -- still less experimental data -- 
at his disposal to pursue the discussion.

Unfortunately, his topological comments remained unnoticed, and Friedmann could never gain the
satisfaction to see any of his theoretical models confronted to cosmological observations : he died
prematurely in 1925 after an ascent on a balloon (as he was also a meteorologist). 
As the cosmologists 
of the first half of the twentieth century had no experimental means at their disposal to measure the topology of the Universe, 
the vast majority of them lost all interest in the question. But in 1971, George Ellis published an important article \cite{Ell71}
taking stock of recent mathematical developments concerning the classification of
spaces and their possible application to cosmology. A revival of
interest for multi--connected cosmologies ensued, under the lead of theorists like the Russians D.
Sokoloff  and A. Starobinski \cite{Sok}. An
observational program was started up in the Soviet Union, under the direction
of Victor Shvartsman. With the six meter diameter telescope newly installed at
Zelenchuk, in the Caucasus, the phantom sources of which Friedmann had spoken
in 1924, meaning multiple images of the same galaxy, were sought. All these
tests failed: no ghost image of the Milky Way or of a nearby galaxy cluster was
recognized \cite{SS}. This negative result allowed for the fixing of some constraints on the
minimal size of a multi--connected space, but it hardly encouraged researchers
to pursue this type of investigation. Interest again subsided.

Next I started to study the subject in the early 1990's, and my talk at the 1993 Friedmann Conference  was
aimed to renew interest at the international level for this manifestly neglected but fundamental question, calling
upon the joint competences of mathematicians, cosmological theorists, and astronomers. This was not yet the real restart, for that I had to publish, along with my recruited ``three musqueteers'' M. Lachi\`eze-Rey, R. Lehoucq and J.-P. Uzan (to which J. Weeks and A. Riazuelo joined a little bit later), a series of more technical articles in which we indicated new research tracks, such as the method of  ``cosmic crystallography'' \cite{bob} -- see also \cite{Rouklum}.   
The present contribution is a personal account summarizing the spectacular development of cosmic topology that followed, due to outstanding contributions by many researchers all around the world, and, like in the celebrated novel by Alexandre Dumas, to tell what is going on  ``Twenty Years After''. Given the fact that during the two last decades, more than one hundred excellent papers on cosmic topology have been published, I apologize for not quoting all the references that could not find place in this short presentation.  
 
To conclude this introduction I want to recall how my meeting with Andrey Grib was fruitful : sharing a mutual interest for the early history of relativistic cosmology, we undertook the first translation (in English and in French) of the semi-popular book that Friedmann devoted to general relativity and cosmology \cite{Fri23}, eventually published in France in 1996 along with a large introduction, technical annotations and historical comments \cite{jplgrib96}.

\section{Topology and Cosmology}
\label{sec:2}

It is presently believed that our Universe is correctly described at large scale by a Friedmann-Lema\^{\i}tre (hereafter FL) model. The FL models are homogeneous and isotropic solutions of Einstein's equations, of which the spatial sections have constant curvature. The FL models fall into 3 general classes, according to the sign of their spatial curvature $k= -1, 0, +1$. The spacetime manifold is described by the line element $ds^2 = c^2dt^2 - R^2(t)d\sigma^2$, where $d\sigma^2 = d\chi^2 + {S_k}^2(\chi) (d\theta^2+\sin^2{\theta}d\varphi^2)$ is the metric of a 3--dimensional homogeneous manifold, flat [$k=0$] or with curvature [$k \pm 1$]. The function $S_k(\chi)$ is defined as $\sinh(\chi)$ if $k= -1$, $\chi$ if $k=0$, $\sin(\chi)$ if $k=1$; $R(t)$ is the scale factor, chosen equal to the spatial curvature radius for non flat models.

In most studies, the spatial topology is assumed to be that of the corresponding simply-connected space: the hypersphere, Euclidean space or the 3D-hyperboloid, the first being finite and the other two infinite. However, there is no particular reason for space to have a simply-connected topology. In any case, general relativity says nothing on this subject : to the metric element given above there are several, if not an infinite number, of compatible topologies, and thus of possible models for the physical Universe. For example, the hypertorus $\bf{T}^3$ and the usual Euclidean space $\bf{E}^3$ are locally identical, and relativistic cosmological models describe them with the same FL equations, even though the former is finite and the latter infinite. Only the boundary conditions on the spatial coordinates are changed. Thus the multi--connected cosmological models share exactly the same kinematics and dynamics as the corresponding simply-connected ones -- for instance, the time evolutions of the scale factor $R(t)$ are identical. 

In FL models, the curvature of physical space depends on the way the total energy density of the Universe may counterbalance the kinetic energy of the expanding space. The normalized density parameter $\Omega_0$, defined as the ratio of the actual energy density to the critical value that an Euclidean space would require, characterizes the present-day contents (matter, radiation and all forms of energy) of the Universe. If $\Omega_0$ is greater than 1, then space curvature is positive and geometry is spherical; if $\Omega_0$ is smaller than 1 the curvature is negative and geometry is hyperbolic; eventually $\Omega_0$ is strictly equal to 1 and space is locally Euclidean. 

The next question about the shape of the Universe is to know whether space is finite or infinite - equivalent to know whether space contains a finite or an infinite amount of matter-energy, since the usual assumption of homogeneity implies a uniform distribution of matter and energy through space.

In multi-connected topologies, one can consider the observable space as part of an idealized ``covering space'', namely the simply-connected homogeneous  spatial section of spacetime, filled with copies of physical space - called the ``fundamental domain''.  The latter is a convex polyhedron  which is finite in at least one direction, and is repeated in the covering space through the transformations of a  ``holonomy group''.  Since Euclidean and hyperbolic covering spaces are infinite, they are tiled by infinite copies of the fundamental domain, whereas in spherical geometry, which is always of finite volume, there are a finite number of copies.  As physical fields are repeated in every copy, they can be can defined in the covering space with periodic boundary conditions. 
Note that topological identifications always break isotropy, and in some cases also the global homogeneity. By chance the most studied spherical and Euclidean spaces  stay homogeneous, but multi-connected hyperbolic  spaces don't.

The multi--connected Euclidean spaces are characterized by their fundamental domains and their holonomy groups. The fundamental domains are either a finite or infinite parallelepiped, or a prism with a hexagonal base, corresponding to the two ways of tiling  Euclidean space. The various combinations generate seventeen multi--connected Euclidean spaces (for an exhaustive study, see \cite{Ria04a}), the simplest of which being the hypertorus $\bf{T}^3$.
Seven of these spaces (called slabs and chimneys) are of infinite volume. The ten other are of finite volume, six of them being orientable hypertori. The latter present a particular interest for cosmology, since they could perfectly model the spatial part of the so-called ``flat'' universe models, which are favored by (too) many cosmologists on the basis of observational data and strict belief in standard inflation. Indeed theoretical arguments coming from quantum cosmology and Wheeler-DeWitt equation favor the flat case : Linde \cite{Lin04} calculated that the spontaneous creation of compact flat universes is much more probable than for other models. 

In a space of  non-zero curvature, contrarily to the Eucidean case, topological compactification cannot be done at an arbitrary scale, due to the presence of a natural length scale : the curvature radius. The topological properties  are now linked to some local features, such as the energy density  $\Omega_0$.

All spherical spaces are finite whatever their topology. The reason is that  the covering space -- the simply-connected hypersphere  $\bf{S}^3$ --  is itself compact;  thus, if one identifies the points of the hypersphere by holonomies belonging to one of the cyclic groups of order $p$, the dihedral groups of order $2m$ or the three binary polyhedral groups which preserve the shapes of the regular polyhedra, the resulting spaces are spherical, multi--connected and compact.  For an exhaustive classification, see \cite{Gaus01}. There is a countable infinity of these, because of the integers $p$ and $m$ which parametrize the cyclic and dihedral groups; but 
there is only a finite set of ``well-proportioned'' topologies, i.e. those in which the fundamental domain is finite with roughly comparable lengths in all directions,. As shown in  \cite{Wee04}, they are of a particular interest for cosmology.

As a now celebrated example, let us mention the Poincar\'e Dodecahedral Space (hereafter PDS), obtained by identifying the opposite pentagonal faces of a regular spherical dodecahedron after rotating by $36^{\circ}$ in the clockwise direction around the axis orthogonal to the face. Its volume is 120 times smaller than that of the hypersphere with the same radius of curvature. After my team ~\cite{Lum03} proposed, in 2003, this space as a plausible candidate for explaining the observed power spectrum of the Cosmic Microwave Background (see below), the mathematical properties of PDS were extensively studied \cite{pds}. It provides an interesting example of how cosmological considerations may impulse new developments in pure mathematics.

Hyperbolic manifolds are less well understood than the other homogeneous spaces. However,
according to the pioneering work of Thurston \cite{Thu82},  almost all manifolds can be endowed
with a hyperbolic structure. Three-dimensional hyperbolic manifolds have a remarkable property that links topology and geometry : the rigidity theorem implies that geometrical quantities such as the volume, the length of its shortest closed geodesics, etc., are topological invariants.  This suggested the idea of using the volumes to classify the compact hyperbolic space forms.  Such volumes are
bounded below by $V = 0.94271$ (in units of the curvature radius), which correspond to the Weeks manifold  \cite{Wee85}. The publicly available program SnapPea (http://www.geometrygames.org/SnapPea/) is specially useful to unveil the rich structure of compact hyperbolic manifolds. Several millions of them with volumes less than 10 could be calculated.  In particular, horned topologies with a finite volume have been invoked to  explain the suppression of  the lower multipoles in the CMB anisotropy \cite{ALST04}.
Quite recently, it was shown that present-day observational constraints on the curvature of space, as well as large-scale anomalies observed in the CMB power spectrum, are  compatible with a marginally hyperbolic space \cite{Lid13}. 

To conclude this geometrical section, note also that so far, the main focus was on classical topologies that contain no singularities, but in string theory, the class of allowed topologies is larger, including the so--called orbifolds \cite{Rat13}.

\section{Observational methods}
\label{sec:3}

From an astronomical point of view, it is necessary to distinguish between the ``observable universe'', which is the interior of a sphere centered on the observer and whose radius is that of the cosmological horizon (roughly the radius of the last scattering surface, presently estimated at 14.4 Gpc), and the physical space. There are only three logical possibilities. 

First, the physical space is infinite -- like for instance the simply-connected Euclidean space. In this case, the observable universe is an infinitesimal patch of the full universe and, although it has long been the preferred model of many cosmologists, this is not a testable hypothesis. 

Second, physical space is finite (e.g. an hypersphere or a closed multi-connected space), but greater than the observable space. In that case, one easily figures out that if physical space is much greater than the observable one, no signature of its finitude will show in the observable data. But if space is not too large, or if space is not globally homogeneous (as is permitted in many space models with multi-connected topology) and if the observer occupies a special position, some imprints of the space finitude could be observable. 

Third, physical space is smaller than the observable universe. Such an apparently odd possibility is due to the fact that space can be multi-connected and have a small volume. 
There are a lot of possibilities, whatever the curvature of space. The present observational constraints on the $\Omega_0$ parameter favor a spatial geometry that is either flat or nearly flat, e.g. slightly positively curved (\cite{wmap9} \cite{planck16}). However, even with the curvature severely constrained by cosmological data, there are still  possible multi-connected topologies that support a homogeneous metric with given nearly flat constant curvature. Such small universe models generate multiple images of light sources, in such a way that the hypothesis can be tested by astronomical observations. The smaller the fundamental domain, the easier it is to observe the multiple topological imaging. 
How do the present observational data constrain the possible multi-connectedness of the universe and, more generally, what kinds of tests are conceivable (see \cite{Lum07} for a non-technical book about all the aspects of topology and its applications to cosmology) ?

Different approaches have been proposed for extracting information about the topology of the universe from experimental data. One approach is to use the 3D distribution of astronomical objects such as galaxies, quasars and galaxy clusters : if the universe was finite and small enough, we should be able to see ``all around'' it, because the photons might have crossed it once or more times. In such a case, any observer might identify multiple images of a same light source, although distributed in different directions of the sky and at various redshifts, or to detect specific statistical properties in the distribution of faraway sources. Various methods of {\it cosmic crystallography} have been devised by my team \cite{bob} \cite{ull99} and widely developed by other groups \cite{mrb} \cite{bresil} \cite{fuji}. The main limitation of cosmic crystallography is that the presently available catalogs of observed sources at high redshift are not complete enough to perform convincing tests. 

The other approach uses the 2D cosmic microwave background (CMB) maps (for a review, \cite{Levin02}). 
The last scattering surface which emitted the CMB is the largest possible volume within which the topology of the Universe can be probed. The two most important CMB methods are the analysis of the power spectrum and the circles-in-the-sky tests.

The early Universe was crossed by acoustic waves generated soon after the big bang. Such vibrations left their imprints 380 000 years later as tiny density fluctuations in the primordial plasma. Hot and cold spots in the present-day 2.7 K CMB radiation reveal those density fluctuations and yield a wealth of information about the physical conditions that prevailed in the early Universe, as well as present geometrical properties like space curvature and topology. Density fluctuations may be expressed as combinations of the vibrational modes of space, whose shape can thus be ``heard'' in a unique way. The relative amplitudes of each spherical harmonics determine the power spectrum, which is a signature of the space geometry and of the physical conditions which prevailed at the time of CMB emission. We have calculated the harmonics (precisely the eigenmodes of the Laplace operator) for most of the spherical topologies~\cite{Leh02}, and did the same for all 18 Euclidean spaces~\cite{Ria04a}.

Now the CMB angular power spectrum observed by the WMAP telescope exhibits values of some large-angle multipoles lower than what is predicted by the standard flat $\Lambda$CDM cosmology, as well as other anomalies confirmed by the last Planck telescope results \cite{planck15}, suggesting that long wavelengths are missing. A topological explanation may be because space is not big enough to sustain long wavelengths. As we have shown that some finite multi-connected topologies called ``well-proportioned'' do lower the large-scale fluctuations~\cite{Wee04}, several authors have proposed various small universe models to account for the missing fluctuations. The best fits were obtained with the positively curved 
Poincar\'e Dodecahedral Space~\cite{Lum03} \cite{Cai07} \cite{Rouk08}) and the flat hypertorus~\cite{Aur08} \cite{Bie08}. However other authors~\cite{PhilKog06} \cite{Kunz06} claimed that this was not a likely explanation for the power spectrum anomalies, because to gain all the possible information from the correlations of CMB anisotropies, one has to consider the full covariance matrix rather than just the power spectrum.

A direct observational method to detect topological signatures in CMB maps, called ``circles-in-the-sky'' \cite{CSS98}, uses pairs of circles with the same temperature fluctuation pattern, which are the intersections of the last-scattering surface and the observer's fundamental domain, if the latter is small enough. Pairs of matched circles present identical temperature fluctuations seen in different directions and phases in the CMB map, but with the same radius. The pattern of matched circular pairs varies according to the topology. The PDS model~\cite{Lum03}, which predicts six pairs of antipodal circles with an angular radius comprised between $5^{\circ}$ and $55^{\circ}$ (sensitively depending on the cosmological parameters), has become a particularly good (though disputed) candidate given the WMAP and Planck CMB data.

Many authors have searched for matched circles in simple topologies, using various statistical indicators and massive computer calculations, and they obtained diverse results. 
Some claim that most of non-trivial topologies, including PDS and $\bf{T}^3$ are ruled out, e.g.~\cite{CSS04} \cite{CSS06}. They have searched for antipodal or nearly antipodal pairs of circles in the WMAP map, and found no such circles to obtain the lower limit of the cell size as about 24 Gpc. However this constraint cannot be applied to those spaces whose matched circles can be highly deviated from antipodal. Other authors claim that they have found the hints of multi-connected spaces using improved versions of the circles-in-the-sky method (e.g. \cite{Roukema2008a} \cite{Roukema2008b}, \cite{Aurich2008}. 

Indeed the circles-in-the-sky method suffers from various loopholes. Firstly  it cannot be used if the size of physical space is bigger than the last scattering surface. Secondly it ignores several effects which blur the correlation between matched circles. The two most important ones are the Doppler contribution, which arises from the acoustic motion of the baryon-photon fluid at the time of the CMB emission, and the integrated Sachs-Wolfe contribution, which depends on the whole photon path between the emitting point of the last scattering surface and the observer:  in a multi-connected space, two matched points of the last scattering surface are joined to the observer by two different paths, and thus contribute differently to the measured temperature. There are further degrading effects, like the finite thickness of the last scattering surface, and residuals left over by the substraction of foreground sources, which have their own uncertainties. Thirdly, for generic multi-connected topologies (including the well-proportioned ones), the matched circles are generally not antipodal; moreover, the positions of the matched circles in the sky depend on the observer's position in the fundamental domain~\cite{Ria04b}. The corresponding larger number of degrees of freedom for the circles search in the CMB data generates a dramatic increase of the computer time. The search for matched  circle pairs that are not back-to-back has nevertheless been carried out recently by \cite{Vaudrevange2012}, with no obvious topological signal appearing in the seven-year WMAP data. 

Such an unclear situation implies that all those topological spaces are not definitely ruled out by circles-in-the-sky searches. 
Other methods for experimental detection of non-trivial topologies have thus been proposed and used to analyze the experimental data, such as the multipole vectors~\cite{Bie08} and the likelihood method~\cite{planck26}. 
The most up-to-date study has been done using the 2013 data from Planck telescope~\cite{planck26}. The  circle-in-the-sky searches and the likelihood methods yield no detection of a compact topology with a scale below the diameter of the last scattering surface. But the constraints they derived concern only topologies that predict matching pairs of back-to-back circles; even for the topologies that predict back-to-back circles, the constraints do not apply to those universes for which the orientation of the matched circles is impossible to detect due to partial masking on the sky. 

As a provisory conclusion, the overall topology of the Universe, after a difficult start, has now become one of the major concerns in astronomy and cosmology. Even if particularly simple and elegant models such as the PDS and the hypertorus are eventually ruled out by future experiments, many models of multi-connected spaces will not be eliminated as such. In addition, even if the size of a multi-connected space is larger (not too much) than that of the observable universe, we could all the same discover an imprint in the fossil radiation, even while no pair of circles, much less ghost galaxy images, would remain. The topology of the universe could therefore provide information on what happens outside of the cosmological horizon. This is search for the next decade, but I am not sure to be still  available to tell the story at a 2025 A. Friedmann International Seminar\ldots


\begin{thebibliography}{99.}
  
\bibitem{lum94}
J.-P. Luminet :  \textit{Multi-connected cosmologies} in Proc. 2nd A. Friedmann Int. Conf., Ed. Y. Gnedin, A. Grib and V. Mostepanenko, 
Russian Acad. of Sci. and Friedmann Lab. Pub.  St. Petersburg, Russia (1994), pp.73-90

\bibitem{lalu}
M. Lachi\`eze-Rey, J.-P. Luminet : Phys.\ Rep.\  \textbf{254}, 135 (1995)

\bibitem{Fri24}
A. Friedmann : Zeitschr. Phys.  \textbf{21}, 326 (1924)

\bibitem{Ell71}
G. Ellis : Gen. Rel. Grav. \textbf{2}, 7 (1971).

 \bibitem{Sok}
 D. Sokoloff, A. Starobinskii: Sov.Astron. \textbf{19}, 629 (1975)
 
 \bibitem{SS}
D. Sokoloff, V. Shvartsman : Sov. JETP, \textbf{39}, 196 (1974)
 
 \bibitem{bob}
R. Lehoucq, M. Lachi\`eze-Rey, J.-P. Luminet:  Astron. Astrophys., \textbf{313}, 339 (1996);  
R. Lehoucq, J.-P. Luminet, J.-P. Uzan:  Astron. Astrophys., \textbf{344}, 735 (1999)

\bibitem{Rouklum}
J.-P. Luminet, B. Roukema: \textit{Topology of the Universe : Theory and Observation} in ``Theoretical and Observational Cosmology'', M. Lachi\`eze-Rey (Ed.), Kluwer Acad. Pub. (1999), p.117-157  

\bibitem{Fri23}
A. Friedmann: \textit{Mir kak prostranstvo i vremya} (The Universe as Space and Time), Leningrad: Academia (1923)

\bibitem{jplgrib96}
J.-P. Luminet : \textit{Friedmann, Lema\^{\i}tre : Essais de Cosmologie}, Introd., transl. and notes J.-P. Luminet, A. Grib, Paris : Le Seuil (1997) 

\bibitem{Ria04a} 
A. Riazuelo, J. Weeks, J.-P. Uzan, R. Lehoucq, J.-P. Luminet : Phys Rev. \textbf{D69}, 103518 (2004)

\bibitem{Lin04}
A. Linde : Journal of Cosmology and Astroparticle Physics, \textbf{2004}, 004 (2004)

\bibitem{Gaus01}
E. Gausmann, R. Lehoucq, J.-P. Luminet, J.-P. Uzan, J. Weeks : Class. Quant. Grav. \textbf{18}, 5155 (2001).

\bibitem{Wee04}
J. Weeks, J.- P. Luminet, A. Riazuelo, R. Lehoucq : MNRAS \textbf{352}, 258 (2004)

\bibitem{Lum03}
J.-P. Luminet, J. Weeks, A. Riazuelo, R. Lehoucq, J.-P. Uzan : Nature \textbf{425}, 593 (2003)

\bibitem{pds}
P. Kramer:  J. Math. Phys. A: Math. Gen. \textbf{38}  3517 (2005); M. Lachi\`eze-Rey: Class. Quant. Grav. \textbf{21 (9)}, 2455 (2004); 
M. Lachi\`eze-Rey, J. Weeks:  Journal of Physics A:    Math. Theor. \textbf{41} 295209 (2008)  

\bibitem{Thu82}
W. Thurston : Bull. Am. Math. Soc., \textbf{6}, 357 (1982)

\bibitem{Wee85}
J. Weeks : \textit{Hyperbolic Structures on 3-Manifolds}, Ph.D. thesis, Princeton Univ. (1985) ; J. Weeks : \textit{The Shape of Space}, 2nd ed., New York: Marcel Dekker (2001) 

\bibitem{ALST04}
R. Aurich, S. Lustig, F. Steiner, H. Then : Class.Quant.Grav. \textbf{21}, 4901 (2004)

\bibitem{Lid13}
A.R. Liddle, M. Cortes: Phys. Rev. Lett. \textbf{111}, 111302 (2013))

\bibitem{Rat13}
B.Rathaus, A. Ben-David, N. Itzhaki: [arXiv:1302.7161]

\bibitem{wmap9}
G. Hinshaw et al.  : [arXiv: 1212.5226] (2013)

\bibitem{planck16}
Planck Collaboration XVI: [arXiv:1303.5076] (2013)

\bibitem{Lum07}
J.-P. Luminet : \textit{The Wraparound Universe}, (AK Peters New York 2007) 

\bibitem{ull99}
J.-P. Uzan, R. Lehoucq, J.-P. Luminet : Astron. Astrophys. \textbf{351}, 766 (1999)

\bibitem{mrb}
A. Marecki, B. Roukema, S. Bajtlik: Astron. Astrophys. \textbf{435}, 427 (2005) 

\bibitem{bresil}
A. Bernui, A. Teixeira: [astro-ph/9904180] (1999) ; G. Gomero, A.Teixeira, M. Reboucas, A. Bernui: Int. J. Mod. Phys. D \textbf{11}, 869 (2002).

\bibitem{fuji}
H. Fujii, Y. Yoshii : Astron. Astrophys. \textbf{529}, A121 (2011)

\bibitem{Levin02}
J. Levin : Phys. Rep. \textbf{365}, 251 (2002)

\bibitem{Leh02}
R. Lehoucq, J. Weeks, J.-P. Uzan, E. Gausmann, J.-P. Luminet : Class. Quant. Grav. \textbf{19}, 4683 (2002) 

\bibitem{planck15}
Planck Collaboration XV: [arXiv:1303.]5075 (2013)

\bibitem{Cai07}
S. Caillerie, M. Lachi\`eze-Rey, J.-P. Luminet, R. Lehoucq, A. Riazuelo, J. Weeks : Astron. Astrophys.  \textbf{476} 691 (2007) 

\bibitem{Rouk08}
B. Roukema et al. : Astron. Astrophys. \textbf{486}, 55 (2008)

\bibitem{Aur08}
R. Aurich et al. :  Class. Quant. Grav., \textbf{25}, 125006 (2008)

\bibitem{Bie08}
P. Bielewicz, A. Riazuelo :  MNRAS \textbf{396}, 609 (2009)

\bibitem{PhilKog06}
N. Phillips, A. Kogut : Astrophys. J., \textbf{645}, 820 (2006)

\bibitem{Kunz06}
M. Kunz et al. : Phys. Rev., \textbf{D73}, 023511 (2006) 

\bibitem{CSS98}
N. Cornish, D. Spergel, G. Starkman : Class. Quant. Grav. \textbf{15}, 2657 (1998)

\bibitem{CSS04}
N. Cornish, D. Spergel, G. Starkman, E. Komatsu : Phys.Rev. Lett. \textbf{92}, 201302 (2004)

\bibitem{CSS06}
J. Key, N. Cornish, D. Spergel, G. Starkman : Phys.Rev.\textbf{D75} 084034 (2007) 

\bibitem{Roukema2008a} 
B. Roukema, Z. Bulinski, N. Gaudin : Astron. Astrophys. \textbf{492}, 657 (2008)

\bibitem{Roukema2008b}
B. Roukema, Z. Bulinski, A. Szaniewska, N. Gaudin: Astron. Astrophys. \textbf{486}, 55  (2008) 

\bibitem{Aurich2008}
R. Aurich : Class. Quant. Grav., \textbf{25}, 225017 (2008)

\bibitem{Ria04b} 
A. Riazuelo, J.-P. Uzan, R. Lehoucq, J. Weeks : Phys Rev. \textbf{D69},103514 (2004)

\bibitem{Vaudrevange2012}
P. Vaudrevange, G. Starkman, N. Cornish, D. Spergel: Phys. Rev. D 86:083526 (2012)

\bibitem{planck26}
Planck Collaboration XXVI: [arXiv:1303.5086] (2013)

\end{thebibliography}
\end{document}